# Surface structure and multigap superconductivity of $V_3Si$ (111) revealed by scanning tunneling microscopy

Shuyue Ding[1†], Dongming Zhao[1†], Tianxing Jiang[1], Haitao Wang[1], Donglai Feng[2,1,3,4], Tong Zhang[1,3,4*]

[1] *Department of Physics, State Key Laboratory of Surface Physics and Advanced Material Laboratory, Fudan University, Shanghai 200438, China.*

[2] *School of Future Technology and Department of Physics, University of Science and Technology of China, Hefei, 230026, China*

[3] *Collaborative Innovation Center of Advanced Microstructures, Nanjing 210093, China.*

[4] *Shanghai Research Center for Quantum Sciences, Shanghai 201315, China.*

**Abstract:**

$V_3Si$, a classical silicide superconductor with relatively high $T_C$ (~16 K), is promising for constructing silicon-based superconducting devices and hetero-structures. However, real space characterization on its surfaces and superconducting properties are still limited. Here we report the first low-temperature scanning tunnelling microscopy (STM) study on cleaned $V_3Si$ (111) single crystal surface. We observed a $\sqrt{3}\times\sqrt{3}$ superstructure which displays mirror symmetry between adjacent terraces, indicating the surface is V-terminated and reconstructed. The tunneling spectrum shows full superconducting gap with double pairs of coherence peaks, but has a relatively small gap size with comparing to bulk $T_C$. Impurity induced in-gap state is absent on surface defects but present on introduced magnetic adatoms. Upon applying magnetic field, a hexagonal vortex lattice is visualized. Interestingly, the vortex size is found to be field dependent, and the coherence length measured from single vortex at low field is significantly larger than estimated value from bulk $H_{c2}$. These results reflect $V_3Si$ is a multi-band, *s*- wave superconductor.

**Introduction:**

$V_3Si$ is a typical $A_3B$ compound superconductor with a rather high $T_C$ of ~16K among conventional superconductors [1]. As a superconducting silicide, it is compatible with the well-developed silicon technology [2,3] and is potentially useful for building silicon-based superconducting device. Besides, many transition metal silicides possess interesting properties such as topological nontrivial states [4,5], ferromagnetism and noncollinear magnetic structures [6,7]. This raised possibilities of constructing all-silicide magnetic/ topological/ superconducting heterostructures by using $V_3Si$ as substrate, in which novel electron pairing and quasiparticle bound states (such as Majorana zero mode) may emerge. To realize these fascinating applications, obtaining clean surface of $V_3Si$ and investigating its electronic property would be highly required, but so far the related studies are still very limited.

On the other hand, the bulk superconductivity of $V_3Si$ itself is also of interest. Although

$V_3Si$ is believed to have a fully gapped Fermi surface, it is a multiband system which may have different gap size (or sign) on different bands, and show some unusual behaviors [8-15]. In the specific heat [8], field dependence of reversible magnetization [13], and thermal conductivity measurement [14], $V_3Si$ behaves like a single-band BCS superconductor. Meanwhile, microwave responses of $V_3Si$ show evidence of two-gap superconductivity [9], and similar multi-gap behaviors were also reported in superfluid [10] and optical measurements [11]. However, direct gap measurements such as STM [16,17] and photoemission spectroscopy (PES) [18] so far only observed a single gap in $V_3Si$. Since these studies are performed at $T \geq 4.3$ K (not sufficiently low with comparing to $T_C$) and mainly on the (001) surface, it is still necessary to study the gap structure at lower temperatures and on different surface orientations to reveal possible multigap feature.

In this work, we performed a low-temperature STM study on the surface structure and superconductivity of $V_3Si$ (111) single crystal. We obtained atomically flat surface with a $\sqrt{3}\times\sqrt{3}$ (R30°) superstructure. Interestingly, such surface structure displays opposite chirality between the nearest neighboring terraces, indicating the surface is V-terminated and reconstructed (with reduced V-atom density). The tunneling spectrum displays a fully gapped superconducting state with two pairs of coherence peaks at ±1.36 meV and ±1.90 meV. The gap sizes are smaller than the expected value from standard BCS ratio and bulk Tc. We found the gap disappears around bulk $T_C$ but the gap. Impurity induced bound states are not observed at surface defects but emerged at deposited magnetic atoms. Upon applying magnetic field, hexagonal vortex lattices are observed up to B = 10T. Remarkably, the vortex core size decreases as the field increase, and the coherence length measured at low field is much larger than the estimated value from bulk $H_{C2}$. We show that these behaviors can be well explained by multiband superconductivity: the gap observed in tunneling spectrum is not the one dominates bulk superconductivity. There likely exists a band with larger gap and shorter coherence length, but not detected in tunnelling spectrum due to low tunnelling probability. Our results suggest $V_3Si$ is a multiband $s_{++}$ wave superconductor, and its clean (111) surface with a full gap would be suitable for epitaxial growth of silicide heterostructures.

**Results:**

The experiment was conducted in a low-temperature STM system (Unisoku) with a base temperature of 400 mK. $V_3Si$ (111) single crystal (Mateck GmbH) was cleaned by repeated Argon sputtering and post-annealing at 800°C, and then transferred *in-situ* into the STM chamber under a vacuum of $1\times10^{-10}$ mbar. Chemically etched W tip was used after e-beam cleaning and Au surface treatment. STM images were acquired at constant current mode and tunneling conductance (dI/dV) were carried out with standard lock-in method with a modulation of 741 Hz.

The A15 crystal structure of $V_3Si$ is sketched at Fig. 1(a). Si atoms formed a body-centered-cubic unit cell with a constant of $a = 4.7$ Å, and each surface of the cubic contains two V atoms. Along the [111] direction, there are Si and V planes stacking alternatively with a period of 0.14 nm (Fig. 1(b)). The Si plane has a simple hexagonal lattice with an in-plane constant of $a' = \sqrt{2}\,a$ (≈ 6.7 Å) (Fig. 1(c)). While in the V planes, each three V atoms clustered into a "trimer" and then formed a hexagonal lattice with a constant of $a'$. Here we note that the two nearest neighboring V planes (marked by V(I) and V(II)) are mirror symmetric. They (together with the underneath Si layer) actually have opposite "surface chirality" as they do not match with each other after any in-plane translation/rotation.

Fig. 1(d) shows a typical large-scale topographic image of the $V_3Si$ (111) after cleaning. The surface is atomically flat with regular terraces. A line profile taken across the terraces (Fig. 2(a)) indicates they have a uniform height of 0.14 nm, which corresponds to the distance between two neighboring V or Si planes along [111] (Fig. 1(b)). Fig. 2(g) shows the zoomed-in image taken at relatively high $V_b$ (≳300 mV), in which a hexagonal "superlattice" that consists of trimer-shaped unit is observed. We note the period of this superlattice is $\sqrt{3}a'$ (1.14 nm) rather than $a'$, which indicates there should be surface reconstructions. Fig. 1(e) shows another image taken across two neighboring terraces. It is notable that the trimer-shaped units have opposite orientation between two neighboring terraces. This directly evidences that the surface is V-terminated rather than Si, as the former is expected to have such mirror symmetry while the latter does not (see Fig. 1(c)). There are some defects/impurities on the surface, which appear as adatoms (type A) or vacancies (type B) as marked in Fig. 1(g). The exact origin of these defect is unknown at this stage.

To further reveal the surface structure, Fig. 1(h) shows a higher resolution image ($V_b$ = 500 meV) with superimposed V and Si lattices of $V_3Si$ (111). It is seen that the positions of trimer units match a part of V trimers of the V lattice. However, only 1/3 of the V atoms of an ideal V lattice (solid red spots in Fig. 1(h)) are identified in the STM image and give rise to the $\sqrt{3}\times\sqrt{3}$ superlattice. Further, in the image acquired at a lowered $V_b$ (Fig. 1(i)), we can resolve another hexagonal lattice with a constant just equals to $a'$. Particularly, its position matches the underlying Si lattice by comparing Figs. 1(i,h) with Fig. 1(c). Therefore, we conclude the observed surface with trimer-shaped units is reconstructed V surface. The formation of such reconstruction could be due to that the ideal V surface is charge polarized, thus reduced surface atom density will help to lower the surface energy.

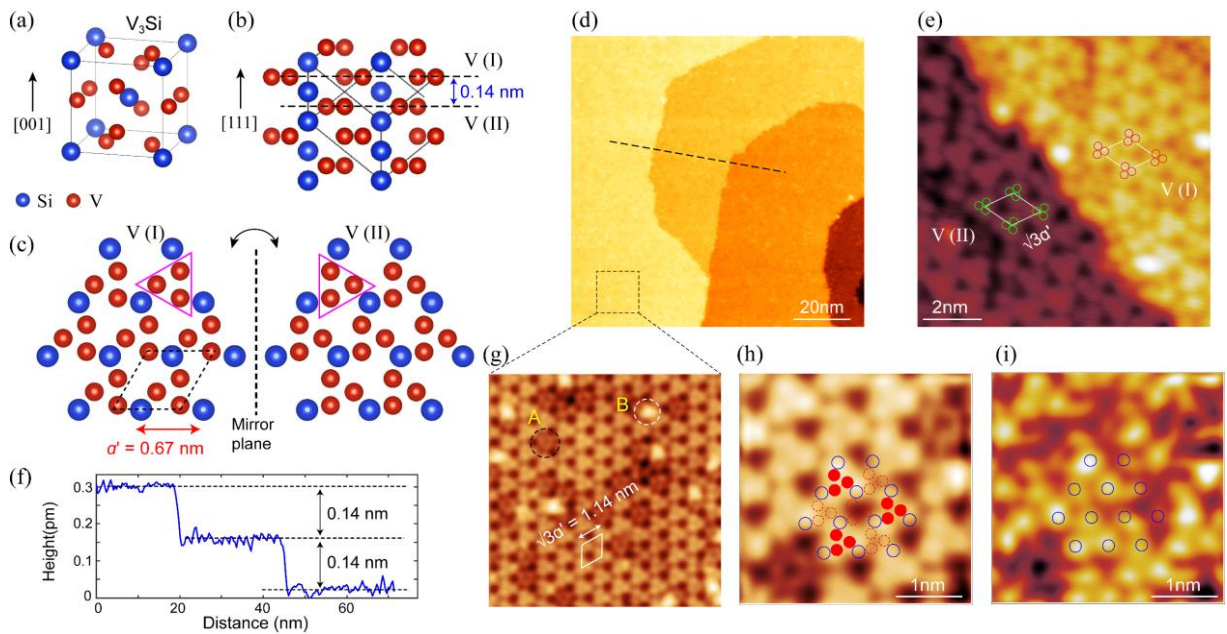

**Fig. 1** | Atomic structure of $V_3Si$ and its STM characterization. **(a)** The A15 cubic structure of $V_3Si$. **(b)** Side view of $V_3Si$ (111) surface. **(c)** Top view of the unreconstructed, V-terminated $V_3Si$ (111) surface (and the

underneath Si layer). There are two kinds of such surface with opposite chirality (V(I) and V(II)), which are mirror symmetric. **(d)** Topographic image of the $V_3Si$ (111) (100×100 nm$^2$, setpoint: $V_b$ = 1V, $I$ = 10pA). **(e)** STM image of two adjacent $V_3Si$ (111) terraces (10×10 nm$^2$). The unit cell of superlattice and the trimer unit on each terrace are indicated. **(f)** Line profile taken along the dashed line in (d). **(g)** Topographic image of a terrace on $V_3Si$ (111) (15×15 nm$^2$, $V_b$ = 950 mV, $I$ = 50pA). The black (white) dash circle indicates typical vacancy (adatom) like surface defects. **(h), (i)** Zoomed-in images (5×5nm$^2$) taken at the same region with $V_b$ = 500mV and 100 mV, respectively. The V and Si lattice are superimposed.

Fig. 2(a) gives a typical large-energy scale dI/dV spectrum (± 1V) of the reconstructed surface. The spectrum shows a metallic behavior with some peak features around 380 meV, -130 meV and -780 meV. We note previous band calculations of $V_3Si$ and A15 superconductors have predicted high DOS peaks near Fermi level which is caused by Von-Hove singularity [15, 19-22]. However such feature is not evidenced here, which may be due to orbital dependent tunneling or the influence of surface reconstruction.

Then we turn to the measurement of superconducting state. Fig. 2(b) shows low-energy dI/dV spectrum (±5 mV) taken on clean surface at T = 400 mK (with optimized energy resolution). A "U" shaped superconducting gap with flat bottom is clearly observed. Notably, there are signatures of two coherence peaks at each gap edge (marked by arrows), which evidenced a multiband superconductivity. We thus used a two-band model to fit the spectrum. The DOS of a single band with isotropic superconducting gap $\Delta$ is given by standard BCS expression:

$$n_s \propto Re[\frac{|E - i\Gamma|}{\sqrt{(E - i\Gamma)^2 - \Delta^2}}]$$

Here $\Gamma$ is a broadening factor due to finite quasiparticle lifetime. The total DOS is then given by: $n_{tot} = \rho_1 n_{s1} + \rho_2 n_{s2}$, where $\rho_1$ and $\rho_2$ are the relative weight of the two bands with superconducting gap size of $\Delta_1$ and $\Delta_2$, respectively. With considering the temperature and system broadening (accounted by a $T_{eff}$ = 1.18 K [23]), the fitting in Fig. 2(b) yields $\Delta_1$= 1.36 meV, $\Delta_2$= 1.90 meV and $\rho_1 = 0.94, \rho_2 = 0.06$. Therefore, the band with smaller gap $\Delta_1$ has overwhelming weight here. We shall note the peak of $\Delta_2$ can only be seen in the spectra with optimized resolution (due to its small weight). In some spectra like the ones in Fig. 2(c, d), there is only one pair of coherence peak but which is more broaden than single BCS gap, thus $\Delta_2$ should be still present (For comparison we show the single gap fit in Fig. 2(b), which clearly deviates the measured spectrum near coherence peaks).

As the $T_C$ of bulk $V_3Si$ is ~16 K, according to the standard BCS ratio of $2\Delta/k_B T_C = 3.53$, the expected gap size for $V_3Si$ is ~2.4 meV. Thus both $\Delta_1$ and $\Delta_2$ are smaller than the expected size. We then measured the temperature dependence of the gap spectrum. As shown in Fig. 2(c), the gap gradually fills up as temperature rises. It became very weak at T = 12.2K, then completely disappear at T ~ 16K. Such quick gap suppression before reaching Tc also imply the observed tunneling gap(s) is not the one dominates bulk superconductivity. This could be due to the superconductivity is weakened at the surface, or there is some (larger) gap residing in a band not detected in tunneling spectrum (similar behavior was observed in typical multiband superconductor $MgB_2$ [24,25]). Our measurements under magnetic field further support this point, as shown later.

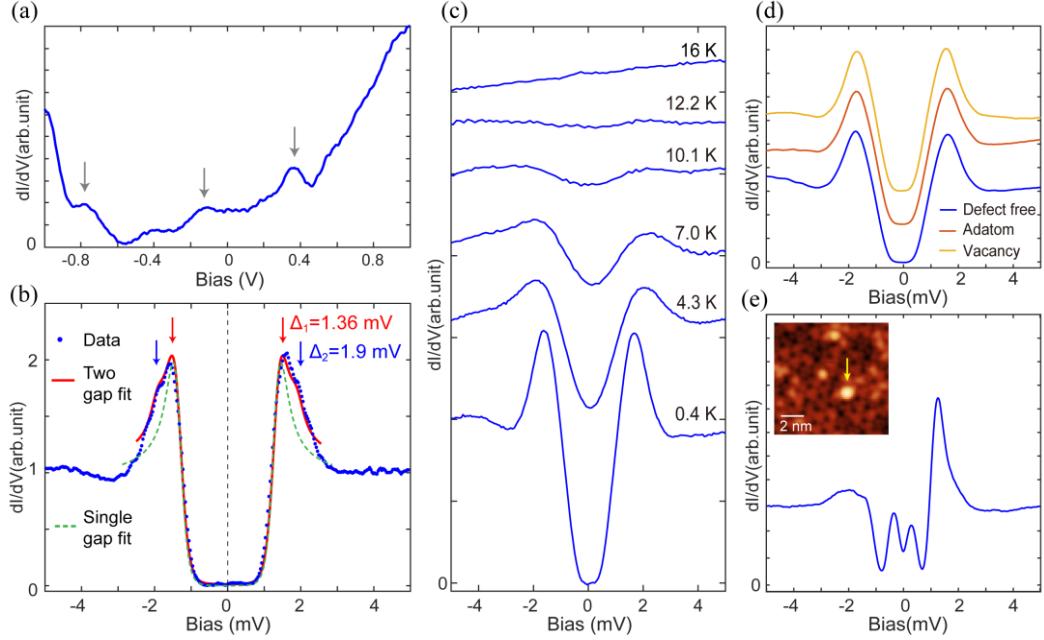

**Fig. 2** (a) Typical large scale dI/dV spectra taken on V$_3$Si (111) surface (setpoint: $V_b$ = 1 V, $I$ = 100 pA). (b). The superconducting gap spectrum of V$_3$Si (111) (setpoint: $V_b$ = 5 mV, $I$ = 100 pA and $T$ = 0.4K). Red curve is the two-band fitting with $\Delta_1$ = 1.36 meV, $\Delta_2$ = 1.90 meV and $T_{eff}$ = 1.18 K. Green curve is the single gap fitting with $\Delta$ = 1.4 meV (c) Temperature dependence of the superconducting gap. (d) Comparison of the gap spectra taken on surface defects and defect-free region. (e) The gap spectrum taken on a deposited Mn atom on V$_3$Si (111) (setpoint: $V_b$ = 5 mV, $I$ = 100pA, T = 400 mK). Inset: STM image around a Mn adatom (marked by the arrow).

Before turn to measurement under magnetic field, we examined the superconducting spectrum on the surface defects, since impurity effect can provide important information on the electron pairing [26]. Fig. 2(d) compares the spectra taken on the adatom-like, vacancy-like defects and defect-free regions. All these spectra display a full superconducting gap without any evidence of in-gap bound states or gap suppression. Assuming the defects are effective scatterers, this observation suggests V$_3$Si is a conventional s-wave superconductor and the adatom/vacancy impurities are non-magnetic. We also intentionally deposited magnetic atoms (Mn) on the surface to investigate the impurity effect. As shown in Fig. 2(e), pronounced in-gap states are observed on Mn atoms, which further supports the s-wave pairing. The impurity effect observed here is consistent with the transport measurement in ref. [15].

V$_3$Si was shown to be a high-$\kappa$ superconductor which supports the vortex state [27], while local measurement on such state is still limited [28]. Here we studied the vortex state under out-of-plane field. Figs. 3(a-g) show the zero-bias conductance (ZBC) maps of a 100×100 nm$^2$ area taken at B = 0.5T ~ 10T. Uniform hexagonal vortex lattices are clearly visualized in these maps, where vortex cores appear as high ZBC region. For the vortices measured at B = 0.5 T (Fig. 3(a)), the distance between vortices (~ 90 nm) is much larger than their lateral size, thus we can consider them as isolated vortices which have weak inter-vortex interaction. Fig. 4(a) shows a series of dI/dV spectra taken across such vortex. It is seen that the gap is gradually suppressed as approaching the core center, which confirms its superconducting origin. Meanwhile, there is no obvious vortex bound state observed and the spectrum at core center is rather featureless. The absence of vortex bound state is usually caused by strong dephasing scattering of quasiparticles [29,30]. Since bulk $V_3Si$ was reported to be in the clean limit [27], the surface impurity scattering likely plays a dominant role here.

As the gap is closed at vortex center and no bound states present, we can estimate the coherence length $\xi_{GL}$ by fitting the gap variation with Ginzburg-Landau (GL) expression of order parameter at N-S boundary:

$$f(r) = \tanh(r/\sqrt{2}\xi_{GL}) \propto 1 - \sigma(r,0) \qquad \text{Eq. (1)}$$

Here $\sigma(r,0)$ is the normalized ZBC at distance $r$ to the vortex center, then $1 - \sigma(r,0)$ is proportional to the normalized gap size. Fig. 4(h) shows the ZBC profile of a vortex in Fig. 4(a) and the fitting using Eq. 1 yields $\xi_{GL}$ = 6.27 nm. It is known that the coherence length also links to the upper critical field via $H_{c2} = \phi_0/2\pi\xi^2$. The bulk $H_{c2}$ of $V_3Si$ was reported to be 22 T [27] which will give a $\xi(H_{c2})$ of 3.8 nm. However, this value is notably smaller than the $\xi_{GL}$ obtained from fitting isolated vortex. Since $\xi$ is inversely related to $\Delta$ ($\xi \propto hv_F/\Delta$), and the observed gap in tunnelling spectrum (Fig. 2(b)) is smaller than the expected value of BCS ratio, the enlarged $\xi_{GL}$ with comparing to $\xi(H_{c2})$ is expected. This also indicates the tunneling gap is not the one dominates bulk $H_{c2}$.

Another important feature in Figs. 3(a~g) is that as the field increases, the lateral size of the vortex core seems to "shrink". Particularly, the vortex lattice is still observable at B = 10T, which exceeds the "upper critical field" of 8.1T determined by $H'_{c2} = \phi_0/2\pi\xi_{GL}^2$ and $\xi_{GL}$ = 6.27 nm. These behaviors are unexpected for single-gap superconductors, while they are evidence of substantial vortex overlapping (or say field dependent coherence length) which have been observed in multi-band superconductors, such as $MgB_2$ and $NbSe_2$ [31-33], and reduced vortex size was also observed in the μSR measurement of $V_3Si$ [34]. Therefore, our measurements point to the existence of multi-band superconductivity. We note a recent band calculation [15] suggests $V_3Si$ has at least four bands crossed $E_F$ with different orbital characters. These bands could have different gap size and tunneling probabilities. The small tunneling gap size and enlarged $\xi_{GL}$ indicate the observed gap is not the one dominates bulk superconductivity. A possible explanation is that there exist a "hidden" band with larger superconducting gap and shorter coherence length, but not detected in tunneling spectrum due to low tunneling probability. This assumption is consistent with the absence of high DOS feature at $E_F$ in the large-scale spectrum (Fig. 2(a)), as the calculation predicted there is band(s) with high DOS near $E_F$ which play a major role in the superconductivity of $V_3Si$ [22].

On the other hand, the reconstructed V-surface of $V_3Si$ may also play a role. The non-stoichiometry (missing V atoms) could weaken the superconductivity near surface and affect the tunneling to bulk band, which results a smaller tunneling gap with enlarged coherence length. The persistence of vortex at B=10T then can be explained by the couplings between surface and bulk superconducting gap (the latter plays a similar role as the "hidden" gap mentioned above). At this stage we cannot pin down such surface effect on superconductivity. Nonetheless, even the surface effect exists, the double coherence peaks in tunneling spectrum would still evidence a multi-gap superconductivity.

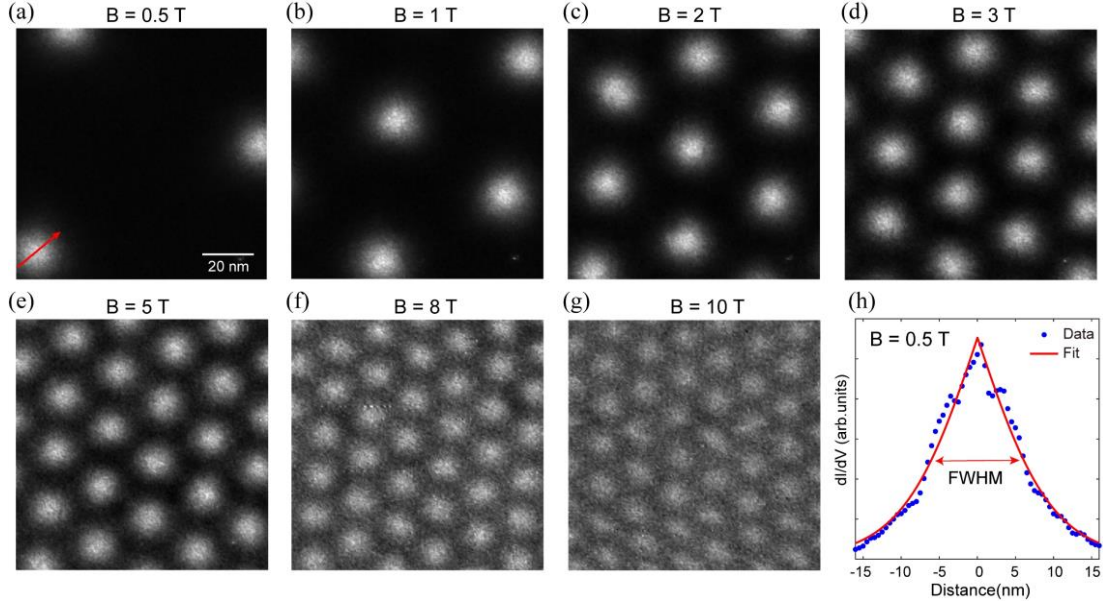

**Fig. 3 | Vortex imaging on V$_3$Si (111) surface. (a-g)**. Zero-bias dI/dV mapping taken at different magnetic field as labeled in each mapping (Scan size: 100×100 nm$^2$, T = 0.4 K, setpoint: $V_b$ = 10 mV, $I$ = 100 pA). **(h)** Line profile of a vortex core in (a) (blue dots) and the fitting by using Eq. (1). and the arrow indicates the FWHM of the vortex core.

At low field (e.g., B = 0.5T), the observed vortices are contributed by the tunneling gap (or surface gap) with small size and larger $\xi_{GL}$. As the field increases, these vortices will phenomenologically overlap and the tunneling gap is suppressed faster than the larger gap of "hidden" band (or bulk gap), thus the apparent vortex size shrinks. In other words, the system will have a field dependent coherence length. To quantitatively describe this behavior, we plot the width (FWHM) of the vortex profile as function of field in Fig. 4(c). We also estimated vortex FWHM by using a simple vortex overlapping model (i.e., arrange the vortices with low-field size into a hexagonal lattice expected for higher fields), the simulated vortex lattices are shown in Fig. 4(e)). The estimated FWHMs fit the measured ones at low field region (Fig. 4(c)), but deviates at higher fields. This is possibly due to finite couplings to the "hidden" gap, which will affect the observed tunneling gap and its coherence length when the field approaches $H'_{c2}$ [32].

Fig. 4(b) shows a series of superconducting gap spectra measured away from vortex core at different fields (taken on the midpoint of three vortices). One can see that besides the expected filling of the gap at increased field, a double-gap feature shows up at B = 8T and 10T. Since our high-resolution spectrum in Fig. 2(b) revealed two gaps at B = 0 ($\Delta_1$ and $\Delta_2$), such double-gap feature likely result from significant reduction of $\Delta_1$ at high fields while $\Delta_2$ keeps almost unchanged, as indicated by the red and black dashed lines in Fig. 4(b) (respectively). It is consistent with that the smaller gap should be suppressed faster when the field increases, but still have non-zero value at $B > H'_{c2}$ due to finite inter-band coupling. $\Delta_2$ is less affected due to its larger size and thus higher critical field. Fig. 4(d) shows the ZBC values extracted from Fig. 4(b) and the estimated values through vortex overlapping simulation (Fig. 4(e)). The estimations also agree with measurements at low fields but deviate at high fields.

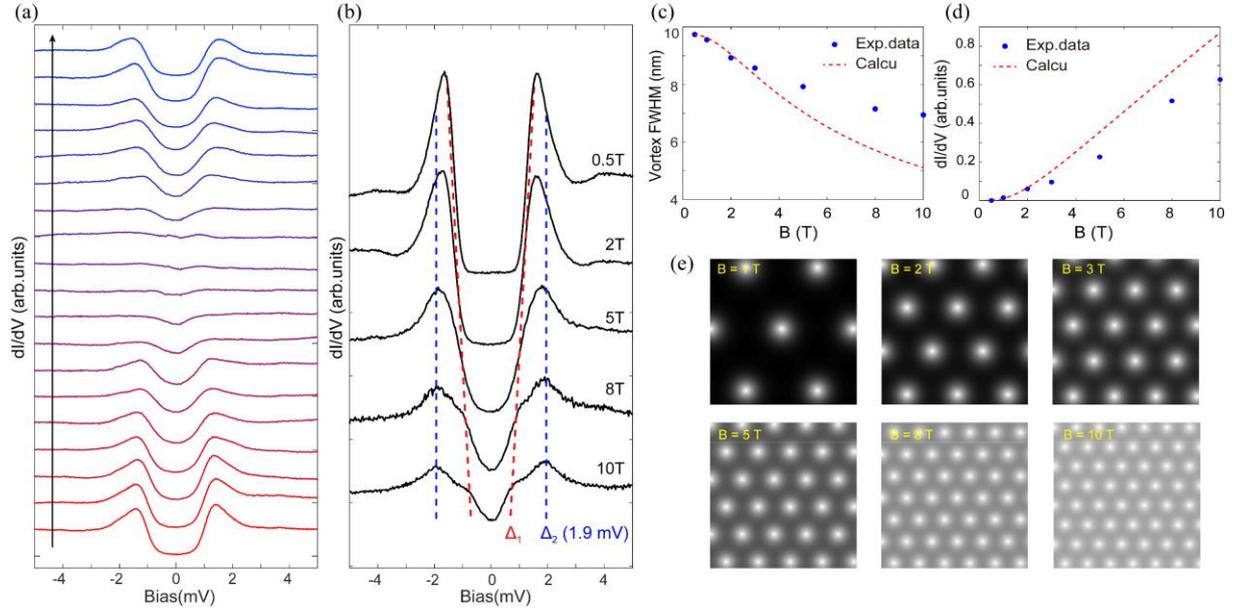

**Fig. 4. (a)** A series dI/dV spectra across the vortex core (taken along the red arrow in Fig. 3(a), setpoint: $V_b$ = 5 mV, $I$ = 100 pA, T=0.4K). **(b)** The field dependence of the dI/dV spectrum taken on midpoint of three vortices at different fields. **(c)** Blue dots: the FWHM of vortex cores extracted from Figs. 3(a-g). Red line: calculated values from panel (e). **(d)** Blue dots: ZBC values extracted from the spectra in panel (b). Red line: calculated ZBC values from. **(e)** Simulated vortex lattice for different fields using vortex overlapping model (see text).

In summary, we performed a systematical study on a cleaned $V_3Si$ (111) single crystal. We observed atomically flat surface with a chiral $\sqrt{3}\times\sqrt{3}$ (R30°) superstructure, indicating the surface is V-terminated with reduced V-atom density. Tunneling spectrum revealed a full superconducting gap which closed around $T_C$ but has a small size with comparing to $T_C$. Impurity effect suggests a conventional pairing state without sign change. Detailed measurement under magnetic field show that the vortex size is field dependent, and the coherence length measured at low field is larger than that expected from bulk $H_{c2}$. These measurements evidence a multi-band, *s*-wave superconductivity in $V_3Si$. The clean $V_3Si$ (111) surface with a full superconducting gap make it as good substrate for fabricating silicon-based superconducting heterostructures.

*Email: tzhang18@fudan.edu.cn
†These authors contributed equally.

**Declarations:**

**Acknowledgements:** Not applicable.

**Funding:** This work is supported by the Innovation Program for Quantum Science and Technology (Grant no. 2021ZD0302800), National Natural Science Foundation of China (Grants Nos. 92065202, 11888101, 11790312, 11961160717, 12225403), Shanghai Municipal Science and Technology Major Project (Grant No. 2019SHZDZX01). Science and Technology Commission of Shanghai Municipality, China (Grant No. 21TQ1400100)

Data availability: All the data and materials relevant to this study are available from the corresponding author upon reasonable request.

Author contributions: S. Ding, D. Zhao, T. Jiang, H. Wang performed the STM measurement. D.L. Feng and T. Zhang coordinated the project. All the authors were engaged in writing and revising the manuscript.

Competing interests: The authors declare no competing interests.

Ethics approval and consent to participate: Not applicable.

Consent for publication: Not applicable.